\documentclass[aps,prb,twocolumn,showpacs,superscriptaddress,longbibliography]{revtex4-2}  
\usepackage{graphicx}  
\usepackage{dcolumn}   
\usepackage{bm}        
\usepackage{amsmath}
\usepackage{amsthm}
\usepackage{amssymb}
\usepackage{bbold} 
\usepackage{mathrsfs}
\usepackage{array}

\usepackage[dvipsnames,usenames]{color}
\usepackage{soul} 
\usepackage{ulem} 
\usepackage{physics}
\usepackage{ulem}

\usepackage{color}
\usepackage[dvipsnames,usenames]{color}

\usepackage{braket}
\usepackage{float}
\usepackage{mhchem}
\usepackage[makeroom]{cancel}
\usepackage{upgreek}
\usepackage{commath}
\usepackage{siunitx}
\usepackage{orcidlink}
\usepackage[titletoc]{appendix}

\usepackage{hyperref}
\hypersetup{colorlinks=true, linkcolor=blue, citecolor=blue, urlcolor=blue}

\begin{document}

\title{The two-dimensional optical Su-Schrieffer-Heeger model: ground state and thermodynamic properties}
\author{Jadson L. Portela e Silva
\orcidlink{0000-0003-3480-4843}}
\affiliation{Instituto de F\'isica, Universidade Federal do Rio de Janeiro Cx.P. 68.528, 21941-972 Rio de Janeiro RJ, Brazil}
\affiliation{Institut f\"ur Theoretische Physik und Astrophysik, Universit\"at W\"urzburg, 97074 W\"urzburg, Germany}
\author{Gabriel Rein \orcidlink{0009-0008-0712-5814}}
\affiliation{Institut f\"ur Theoretische Physik und Astrophysik, Universit\"at W\"urzburg, 97074 W\"urzburg, Germany}
\author{S. dos A. Sousa-Júnior
\orcidlink{0000-0002-4266-3780}}
\affiliation{Department of Physics, University of Houston, 77204, Houston, Texas, USA}
\author{Fakher F. Assaad \orcidlink{0000-0002-3302-9243}}
\affiliation{Institut f\"ur Theoretische Physik und Astrophysik, Universit\"at W\"urzburg, 97074 W\"urzburg, Germany}
\affiliation{W\"urzburg-Dresden Cluster of Excellence ct.qmat, Am Hubland, 97074 W\"urzburg, Germany}
\author{Natanael C. Costa \orcidlink{0000-0003-4285-4672}}
\affiliation{Instituto de F\'isica, Universidade Federal do Rio de Janeiro Cx.P. 68.528, 21941-972 Rio de Janeiro RJ, Brazil}
\affiliation{Institut f\"ur Theoretische Physik und Astrophysik, Universit\"at W\"urzburg, 97074 W\"urzburg, Germany}

\begin{abstract}
We investigate the two-dimensional optical Su-Schrieffer-Heeger (SSH) model, in which the electron hopping amplitude is modulated by the difference between neighboring phonon coordinates. Using sign-problem-free auxiliary-field quantum Monte Carlo simulations, complemented by mean-field analysis, we determine the long-range ordered phases as a function of the electron-phonon coupling and phonon frequency. By examining both adiabatic and antiadiabatic regimes, we reveal the emergence of staggered and armchair valence bond solid (VBS) phases, as well as the O(4) antiferromagnetic phase. In addition, finite-temperature simulations show that the VBS transition occurs at critical temperatures significantly higher than in models with local electron-phonon coupling, consistent with the presence of lighter polarons in the metallic regime. These findings establish the ground-state and finite-temperature phase diagrams of the optical SSH model, which emphasize its similarities and contrasts with other electron-phonon systems.
\end{abstract}

\maketitle
\section{Introduction}
\label{sec:intro}

Many strongly correlated systems ascribe their exotic behaviors primarily to the Coulomb electron-electron interaction.
Examples include the cuprate and iron-pnictide families, whose superconductivity, besides being marked by high transition temperatures, is believed to originate in a spin fluctuation pairing glue as opposed to a conventional electron-phonon coupling\,\cite{Lee2006,Orenstein2000,Sadovskii2008,Keimer2015,Si2016}.
Despite this, for many of these materials, it is believed that accounting for additional effects of interactions with the lattice might be essential for a quantitative understanding\,\cite{Lanzara2001,Shen2002,rosch2005,Lee2006,DeFilippis2009}.
In particular, previous angle-resolved photoemission spectroscopy (ARPES) experiments in cuprates revealed that phonon modes directly affect electron dynamics\,\cite{Lanzara2001}.
Interestingly, recent ARPES measurements for the quasi-1D cuprate Ba$_{2-x}$Sr$_{x}$CuO$_{3 + \delta}$\,\cite{Chen21a} exhibit spectra that are theoretically reproducible only if one adds to the Hubbard model an attractive interaction between nearest neighbor sites\,\cite{Chen21a,Wang21}, whose nature may be due to the coupling to phonon degrees of freedom\,\cite{Scalapino12}.
Similar interesting examples for strong electron-electron and electron-phonon interactions are found in the BSCCO cuprates doped with interstitial oxygens\,\cite{Gomes07,Zeljkovic12,Song19a}.

\begin{figure}[h]
    \includegraphics[scale=0.52]{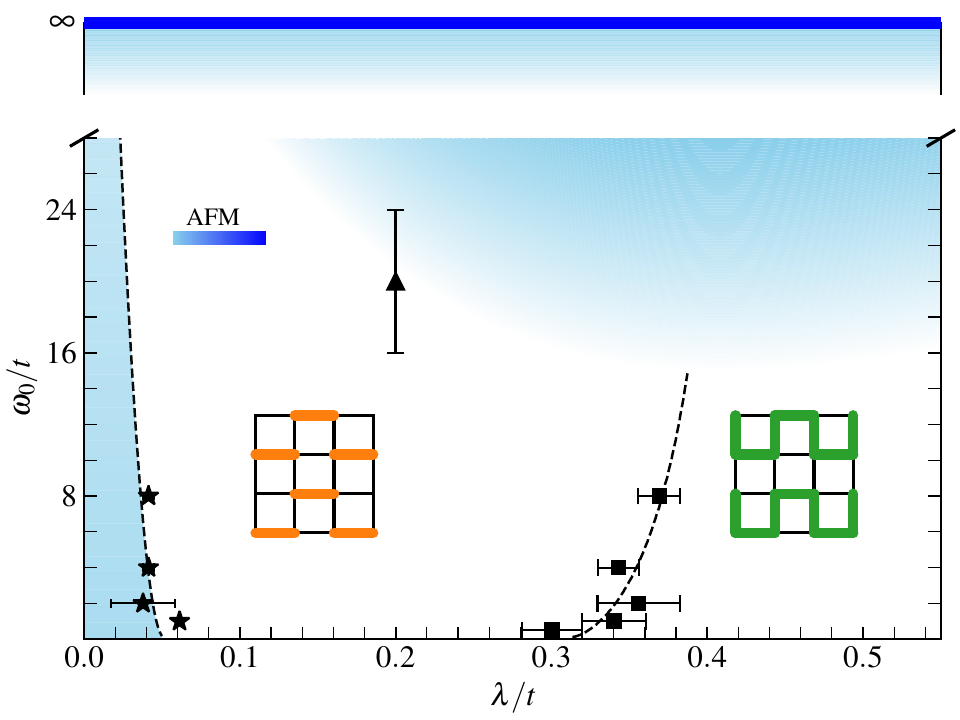}
\caption{The ground state phase diagram of the optical-SSH model for the half-filled square lattice. Black symbols indicate QMC results, while  lines are guides to the eye. Blue shading indicates the parameter region where the AFM/CDW/SC phase occurs. The orange and green patterns indicate the type of VBS ordering.}
\label{fig:phase_diagram}
\end{figure}

Indeed, the interplay between electronic and phononic degrees of freedom may lead to competition between different phases of matter.
However, the emergence of phases and their nature are strongly dependent on the type of fermion-boson coupling.
Within this context, the simplest models to describe electron-phonon phenomena are given by the Holstein\,\cite{Holstein1959} and the Su-Schrieffer-Heeger (SSH) \cite{Su1979,Su1980} model.
The former describes phonons locally coupled to the electronic density, and has been extensively investigated in two-dimensional systems over the past decade, with the emergence of charge-density wave (CDW) or superconductivity (SC)\,\cite{Costa2018,Dee2019,Zhang20192,Chen2019,Costa2020,Feng2020,CohenStead2020,Bradley2021,Costa2020}.
The latter defines phonons coupled to the fermionic kinetic term, modulating the electron hopping.
Despite the vast use of the SSH model for single-particle topological problems, its many-body features is a topic still under debate.
For one-dimensional systems, the model exhibits bond-order wave and charge-density wave, with the emergence of deconfined quantum criticality\,\cite{Fradkin1983,Weber2015,Weber2020,Ferrari2020,Piccioni2025};
a detailed review of the interacting 1D SSH model (and also Holstein model) is presented in Ref.\,\onlinecite{Hohenadler2018} and references therein.
However, its properties for two-dimensional systems are much less clear, even in the absence of Coulomb repulsion.

Before proceeding, we note that the hopping modulation in the SSH model can be implemented in different ways. In the original formulation by Su, Schrieffer, and Heeger\,\cite{Su1979,Su1980}, site displacements of an acoustic mode modify the nearest-neighbor hopping as $t_{ij}=t_0 + g(X_i-X_j)$, which we refer to as the \textit{acoustic-SSH} model. An alternative employs the same hopping modulation while coupling to dispersionless (Einstein) optical phonons; we refer to this as the \textit{optical-SSH} model. Finally, one may couple a bond-local bosonic coordinate $X_{ij}$ to the hopping, such as $t_{ij}=t_0 + g X_{ij}$, again with optical phonons; we refer to this variant as the \textit{bond-SSH} model. Despite the differences on the couplings, these SSH-variants share the same symmetries.

The many-body properties of these SSH models have only recently been investigated in higher dimensions, through unbiased methodologies.
In particular, quantum Monte Carlo (QMC) studies of the two-dimensional bond-SSH model report a transition from a staggered valence-bond solid (VBS) state at large electron-phonon coupling (EPC) to a non-VBS phase at weaker coupling\,\cite{Xing2021}. Subsequent analyses identify this non-VBS phase as being antiferromagnetic (AFM) degenerate with CDW and SC states, as a consequence of an $O(4)$ symmetry of the model\,\cite{goetz2021,Cai2021}. While the available data are consistent with a continuous VBS-AFM transition, it is probably a weak first-order transition, since $C_4$ vortices of the staggered VBS on the square lattice do not carry the topological charge associated with a deconfined critical point\,\cite{Sen2010,Xu2011}. By contrast, other variants of the model exhibit Dirac fermions together with columnar VBS order\,\cite{Goetz2024}, features often used as ingredients for deconfined criticality\,\cite{Goetz2024b,Seifert2024}. Finally, further studies have also introduced a finite (repulsive or attractive) Hubbard-$U$  term\,\cite{Feng2021,Cai2022,Xing2023}. It reduces the symmetry from an $O(4)$ to $SO(4)$, which can be understood as a low-energy $SU(2)_{\rm spin}$ and a high-energy $SU(2)_{\eta}$ for $U>0$ (and vice-versa for the attractive case), thereby lifting the degeneracy between AFM and the CDW/SC channels.

Despite the great efforts to understand the bond-SSH model, the acoustic and optical SSH models have been less explored. Recent QMC studies of the optical-SSH model at half filling on the square lattice indicate that a nonzero critical EPC is required to stabilize a staggered VBS phase\,\cite{Malkaruge2023,Tanjaroon2023,Tanjaroon2025}. For EPC strengths below this threshold, unlike in the bond-SSH case, the data show only weak, short-range AFM correlations, consistent with metallic behavior\,\cite{Tanjaroon2025}. Introducing a local Hubbard interaction promotes long-range AFM order, yielding a phase diagram in which AFM and VBS tendencies compete. A related picture has been reported on the honeycomb lattice at half filling, where a semimetal-to-Kekul\'e VBS transition occurs\,\cite{Malkaruge2024}. The latter behavior is expected, due to the vanishing density of states at the Dirac points on the honeycomb lattice. However, on the square lattice, Fermi surface nesting and the van Hove singularity favor interaction-driven instabilities, thus a metallic phase should be disfavored. Therefore, the possible existence of a state that is neither VBS nor AFM/CDW/SC seems interesting and requires closer examination.

In view of this, we investigate in detail the ground state and thermodynamic properties of the two-dimensional optical-SSH model, using sign-problem-free auxiliary-field quantum Monte Carlo (AFQMC) simulations, complemented by mean-field analysis.
Our main result is summarized in the rich phase diagram of Figure \ref{fig:phase_diagram}, which encompasses staggered and armchair VBS regions, as well as transitions to the AFM/CDW/SC phase, from adiabatic ($\omega_{0}\to 0$) to anti-adiabatic ($\omega_{0}\to \infty$) regimes. We also present a finite-temperature phase diagram in Fig.\,\ref{fig:diagramT}, with critical temperatures considerably higher than in models where phonons couple to the charge density, such as in the Holstein model. These results clarify the recent discussions in the literature and provide a numerical benchmark for this model, whose details are presented below.
The paper is organized as follows: Sec.\,\ref{sec:model} introduces the model and its symmetries, as well as the methods. Section \ref{sec:results} presents the numerical results and discussions, where we determine the emergence of long-range ordered phases. Finally, Sec.\,\ref{sec:conclusions} summarizes our main conclusions.

\section{The optical SSH model}
\label{sec:model}

\subsection{The model}

In this work, we investigate the properties of the optical-SSH model, highlighting its similarities with other electron-phonon models when needed.
The Hamiltonian of the optical-SSH model reads
\begin{align}
\nonumber    \mathcal{H} = & -t \sum_{\langle \textbf{i,j} \rangle,\sigma} K_{\textbf{i,j}}^\sigma - \mu \sum_{\textbf{i},\sigma} n_{\textbf{i},\sigma} \\
\nonumber
& - \sum_{\alpha=x,y}\sum_{\textbf{i},\sigma}  g_{\alpha} K_{\textbf{i,i}+\hat{\alpha}}^\sigma ( Q_{\alpha,\textbf{i}} - Q_{\alpha,\textbf{i} +\hat{\alpha}}  )  \\
& +\sum_{\alpha=x,y}\sum_\textbf{i} \left( \dfrac{P_{\alpha,\textbf{i}}^2}{2M} + \dfrac{M \omega^2_0}{2}Q_{\alpha,\textbf{i}}^2   \right)  . 
\label{Eq:SSHH}
\end{align}
where the sums run over a square lattice, with $\langle \mathbf{i},\mathbf{j} \rangle$ denoting nearest neighbor sites. Here, $$K_{\textbf{i,j}}^\sigma = \left( c^{\dagger}_{\mathbf{i}\sigma} c_{\mathbf{j}\sigma} + \mathrm{H.c.} \right)$$ is the hopping operator, with $c_{\mathbf{i}\sigma}$ ($c^{\dagger}_{\mathbf{i}\sigma}$) being annihilation (creation) operators of electrons on a given site $\mathbf{i}$ and spin $\sigma$.
The first two terms on the right-hand side of Eq.\,\eqref{Eq:SSHH} correspond to the kinetic energy of electrons, and their chemical potential, respectively.
The third term denotes the EPC, with $g_{\alpha}$ defining its strength.
Finally, the fourth term describes the phonon degrees of freedom along \textit{x} and \textit{y}-coordinates, where $P_{\alpha,\textbf{j}}$ and $Q_{\alpha, \textbf{i}}$ ($\alpha = x, y$) are momentum and position operators of local quantum harmonic oscillators with frequency $\omega_0$, respectively.
For simplicity, but without loss of generality, hereafter we define $g_x = g_y = g $, and we set the energy scale in units of the hopping integral $t$. We also set the Boltzmann ($k_{\rm B}$), Planck ($\hbar$), and lattice ($a$) constants as unity.

At this point, it is important to note that, although both the optical and bond-SSH models modulate the kinetic energy, they exhibit distinct differences. The main difference arises from their electron-phonon coupling term: in the optical model, this term depends on the difference of phonon modes between nearest-neighbors sites, resulting in a wavevector dependence with a node at $\mathbf{q} = (0,0)$. By contrast, such a $q$-dependence is absent in the bond version of the model. As demonstrated in this work, this wavevector dependence plays a crucial role in generating qualitative differences between the two models.

\subsection{Symmetries}
 
Despite this difference, these SSH models exhibit the same symmetries. Both the optical and bond-SSH models on a bipartite lattice are invariant under $O(2N)$ symmetry, regardless of whether the phonons are coupled to the bonds or the sites. This invariance holds as long as the hopping matrix elements occur exclusively between sites belonging to different sublattices. In order to demonstrate this, we examine the problem within the Majorana fermions formulation, by defining
\begin{align}
\label{eq:MajoranaA}
  \mathbf{i} \in A:
    \begin{cases}
      \hat{\gamma}_{\mathbf{i},\sigma,1} = ~~~\big( c_{\mathbf{i},\sigma} + c^{\dagger}_{\mathbf{i},\sigma}\big)\\
      \hat{\gamma}_{\mathbf{i},\sigma,2} = -i\big( c_{\mathbf{i},\sigma} - c^{\dagger}_{\mathbf{i},\sigma}\big)\\
    \end{cases}       
\end{align}
and
\begin{align}
\label{eq:MajoranaB}
  \mathbf{i} \in B:
    \begin{cases}
      \hat{\gamma}_{\mathbf{i},\sigma,1} = -i\big( c_{\mathbf{i},\sigma} - c^{\dagger}_{\mathbf{i},\sigma}\big)\\
      \hat{\gamma}_{\mathbf{i},\sigma,2} = -\big( c_{\mathbf{i},\sigma} + c^{\dagger}_{\mathbf{i},\sigma}\big)~,\\
    \end{cases}       
\end{align}
with the Majorana operators satisfying the anti-commutation relations
\begin{align}
    \big\{ \hat{\gamma}_{\mathbf{i},\sigma,\alpha}, \hat{\gamma}_{\mathbf{j},\sigma^{\prime},\alpha^{\prime}} \big\}  = 2 \delta_{\mathbf{i},\mathbf{j}}\delta_{\sigma,\sigma^{\prime}} \delta_{\alpha,\alpha^{\prime}}.
\end{align}
From Eqs.\,\eqref{eq:MajoranaA} and \eqref{eq:MajoranaB}, we obtain
\begin{align}
\label{Eq:majoranaKij}
\nonumber
\sum_{\sigma} \hat{K}^{\sigma}_{\mathbf{i},\mathbf{j}} = & \sum_{\sigma} \big( c^{\dagger}_{\mathbf{i},\sigma} c_{\mathbf{j},\sigma} + {\rm H.c.} \big) \\
\nonumber
= & \frac{1}{4} \sum_{\sigma} \bigg[ \big(\hat{\gamma}_{\mathbf{i},\sigma,1} - i\hat{\gamma}_{\mathbf{i},\sigma,1} \big) \big(i\hat{\gamma}_{\mathbf{j},\sigma,2} - \hat{\gamma}_{\mathbf{j},\sigma,2} \big) + {\rm H.c.} \bigg] \\
= & \frac{i}{2} \sum_{\sigma} \sum_{\alpha =1}^{2} \hat{\gamma}_{\mathbf{i},\sigma,\alpha} \hat{\gamma}_{\mathbf{j},\sigma,\alpha}~.
\end{align}
We recall that $\hat{K}^{\sigma}_{\mathbf{i},\mathbf{j}}$ is invariant under $U(1)$ transformation, i.e.~by performing $c_{\mathbf{i},\sigma} \to e^{i\theta} c_{\mathbf{i},\sigma}$.
In the Majorana notation, the $U(1)$ transformation corresponds to
\begin{equation}
\label{Eq:O(2N)}
\begin{pmatrix}
\hat{\gamma}_{\mathbf{j},\sigma,1} \\
\hat{\gamma}_{\mathbf{j},\sigma,2}
\end{pmatrix}
    \to
    \begin{pmatrix}
\cos\theta & -\sin\theta\\
\sin\theta & \cos\theta
\end{pmatrix}
\begin{pmatrix}
\hat{\gamma}_{\mathbf{j},\sigma,1} \\
\hat{\gamma}_{\mathbf{j},\sigma,2}~,
\end{pmatrix}
\end{equation}
which leaves Eq.\,\eqref{Eq:majoranaKij} invariant.
As Eq.\,\eqref{Eq:O(2N)} corresponds to the $O(2N)$ rotation group in the Majorana fermion space, with $N=2$, then it leads to an $O(4)$ symmetry.

Therefore, as any generalized SSH Hamiltonian may be written as
\begin{align}
\nonumber
    \mathcal{H} = & -t\sum_{\langle\mathbf{i},\mathbf{j}\rangle \sigma} \hat{K}^{\sigma}_{\mathbf{i},\mathbf{j}} 
    + g \sum_{\langle\mathbf{i},\mathbf{j}\rangle \sigma} F(\hat{X}_{\mathbf{i}},\hat{X}_{\mathbf{j}}) \hat{K}^{\sigma}_{\mathbf{i},\mathbf{j}} + \mathcal{H}_{ph} \\
    = & -\sum_{\langle\mathbf{i},\mathbf{j}\rangle \sigma} \big[t - g F(\hat{X}_{\mathbf{i}},\hat{X}_{\mathbf{j}}) \big] \hat{K}^{\sigma}_{\mathbf{i},\mathbf{j}} + \mathcal{H}_{ph}
\end{align}
with $F(\hat{X}_{\mathbf{i}},\hat{X}_{\mathbf{j}})$ being an arbitrary function, in the Majorana formulation it reads
\begin{align}
\nonumber    \mathcal{H}  = & -\frac{i}{2} \sum_{\langle\mathbf{i},\mathbf{j}\rangle} \big[t - g F(\hat{X}_{\mathbf{i}},\hat{X}_{\mathbf{j}}) \big]  \sum_{\sigma} \sum_{\alpha =1}^{2} \hat{\gamma}_{\mathbf{i},\sigma,\alpha} \hat{\gamma}_{\mathbf{j},\sigma,\alpha} \\ & + \mathcal{H}_{ph}~,
\end{align}
which must be invariant under $O(4)$ symmetry if sublattice symmetry is maintained. Here, $\mathcal{H}_{ph}$ denotes the free phonon term, which can be acoustic or optical.

\subsection{The anti-adiabatic limit}

We also investigate the properties of the Hamiltonian given by Eq.\,\eqref{Eq:SSHH} in the anti-adiabatic limit, i.e., $\omega_0 \to \infty$. In this limit, a purely fermionic description becomes possible by taking $M \to 0$, while maintaining $k = M \omega_{0}^2$ finite. Starting from the action 
\begin{align}
    \nonumber S = S_0 + \int_{0}^{\beta}d\tau \sum_{\mathbf{i},\alpha}\Big[& (Q_{\alpha,\mathbf{i}} -Q_{\alpha,\mathbf{i}+\hat{\alpha}})K_{\mathbf{i},\mathbf{i}+\hat{\alpha}} \\ &+\frac{M\dot{Q}_{\alpha,\mathbf{i}}^2}{2}+\frac{k}{2}Q_{\alpha,\mathbf{i}}^2\Big]~,
    \end{align}
by rearranging the terms and taking $M\to0$, we obtain
\begin{align}
    S_0 + \int_{0}^{\beta}d\tau \sum_{\mathbf{i},\alpha}\Big[Q_{\alpha,\mathbf{i}}(K_{\mathbf{i},\mathbf{i}+\hat{\alpha}}-K_{\mathbf{i}-\hat{\alpha},\mathbf{i}})+\frac{k}{2}Q_{\alpha,\mathbf{i}}^2\Big].
\end{align}
This action has a form amenable to carry out a Gaussian integration, leading to the Hamiltonian
\begin{align}\label{eq:SSH_antiadiabatic}
    \mathcal{H}_{\infty} = -t  \sum_{\langle \mathbf{i},\mathbf{j} \rangle,\sigma} K_{\mathbf{i},\mathbf{j}}^\sigma - \lambda \sum_{\mathbf{i},\alpha} \Big[ \sum_{\sigma} \Big(K_{\mathbf{i},\mathbf{i}+\hat{\alpha}}^{\sigma} - K_{\mathbf{i}-\hat{\alpha},\mathbf{i}}^{\sigma}\Big)\Big]^2
\end{align}
with 
\begin{equation}
\lambda = \frac{g^2}{2 k } = \frac{g^2}{2 M \omega_{0}^{2} }.  
\end{equation}
Hereafter, $\lambda/t$ will be our dimensionless EPC parameter.

\subsection{The methods}

The physical properties of the finite frequencies and anti-adiabatic Hamiltonians are investigated using a finite-temperature AFQMC methodology\,\cite{Blankenbecler81,Hirsch83,Hirsch85,Kawashima2002,rrds2003,scalettar89}. Both cases lead to sign-free AFQMC approach. Specifically, for Eq.\,\eqref{Eq:SSHH}, we employ Langevin dynamics to the phonon modes, while the anti-adiabatic case is handled using the ALF-implementation of the  AFQMC, as described in Ref.\,\onlinecite{ALFcol}. Detailed information about the Langevin implementation may be found in Refs.\,\onlinecite{sorella2017,batrouni2019,goetz2021}.

Here, we investigate the occurrence of a VBS state by analyzing the response of the bond-bond correlation functions, $\langle K_{\mathbf{i,j}}^\sigma K_{\mathbf{p,q}}^{\sigma^{\prime}}  \rangle $, and their Fourier transform for bonds along \textit{x} or \textit{y} directions, i.e.~the bond structure factors
\begin{align}\label{Eq:SBx}
S_{\rm B\,(x)}(\textbf{q}) = \frac{1}{N} \sum_{\textbf{i}, \textbf{j}} e^{i\textbf{q} \cdot (\textbf{i}- \textbf{j})} \langle K_{\textbf{i,i}+\hat{x}}^\sigma K_{\textbf{j,j}+\hat{x}}^\sigma \rangle ~,  
\end{align}
and
\begin{align} \label{Eq:SBy}
S_{\rm B\,(y)}(\textbf{q}) = \frac{1}{N} \sum_{\textbf{i}, \textbf{j}} e^{i\textbf{q} \cdot (\textbf{i}- \textbf{j})} \langle K_{\textbf{i,i}+\hat{y}}^\sigma K_{\textbf{j,j}+\hat{y}}^\sigma \rangle ~,  
\end{align}
with $N = L \times L$ being the number of sites.
Similarly, we examine the emergence of an AFM phase by analyzing the spin structure factor
\begin{align}
S_{\rm S} (\textbf{q}) = \frac{1}{N} \sum_{\textbf{i}, \textbf{j}} e^{-i\textbf{q} \cdot (\textbf{i}- \textbf{j})} \langle S^{z}_{\mathbf{i}} S^{z}_{\mathbf{j}} \rangle ~,  
\end{align}
where $S^{z}_{\mathbf{i}} = (n_{\mathbf{i} \uparrow} - n_{\mathbf{i} \downarrow})$ is the \textit{z}-component of the spin operator.
The emergence of long-range order is probed through finite-size scaling analysis of these quantities.

We also employ a static mean-field approach, in which we assume a permanent distortion that breaks the translation symmetry of the lattice. This is effectively done by neglecting the kinetic term of the phonon fields, which corresponds to the limit $\omega_0 \to 0$, while keeping $M \omega^{2}_0 \equiv k$ finite.
In this limit, the Hamiltonian in Eq.\,\eqref{Eq:SSHH} becomes
\begin{align}
\nonumber   \mathcal{H}_{\rm MFT} = & - \sum_{\alpha=x,y}\sum_{\textbf{i},\sigma} \left[ t + g \left( \langle Q_{\alpha,\textbf{i}} \rangle - \langle Q_{\alpha,\textbf{i} +\hat{\alpha}} \rangle  \right) \right] K_{\textbf{i,i}+\hat{\alpha}}^\sigma  \\
& - \mu \sum_{\textbf{i},\sigma} n_{\textbf{i},\sigma} + \dfrac{M \omega^2_0}{2} \sum_{\alpha=x,y}\sum_\textbf{i} \langle Q_{\alpha,\textbf{i}} \rangle^2 ,
\label{Eq:MFTSSHH}
\end{align}
which is quadratic in the electronic creation and annihilation operators, with the phonon coordinates replaced by their expectation values, $\langle Q_{\mathbf{i},\alpha} \rangle$.
A variational treatment is then appropriate, and leads to the exact ground state, once the ans\"atz is correct. We determine $ \langle Q_{\mathbf{i},\alpha} \rangle$ by self-consistently minimizing the Helmholtz free energy obtained from diagonalizing $\mathcal{H}_{\rm MFT}$.

\begin{figure}[t]
\includegraphics[scale=0.5]{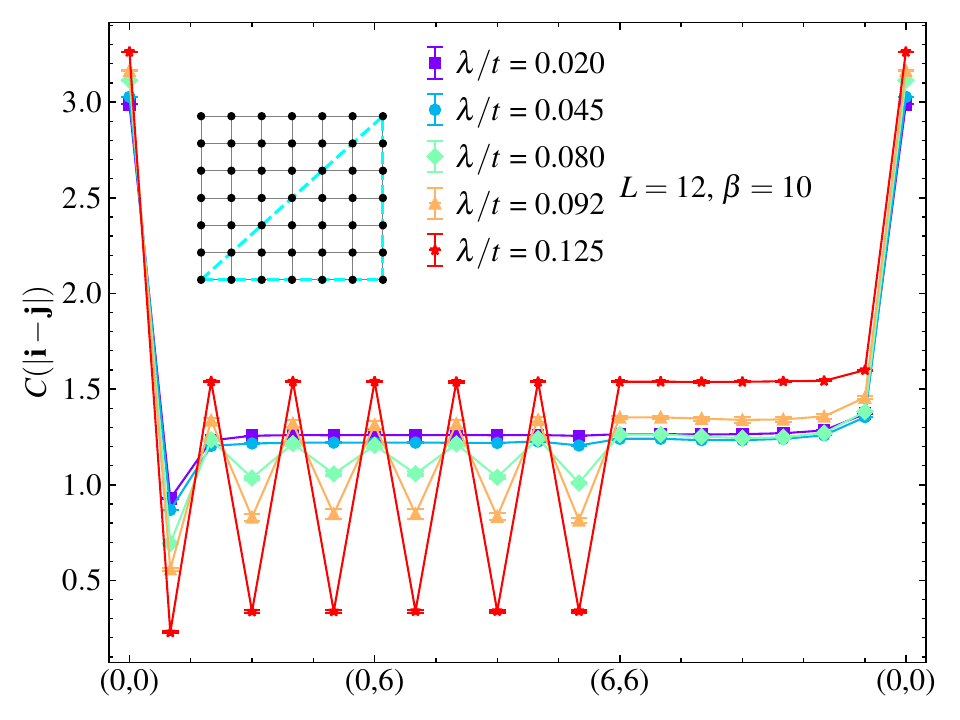}
\caption{Bond-bond correlation function $C(r)$ as a function of the distance $r=|\mathbf{i}-\mathbf{j}|$ between two given sites along high-symmetry directions of the square lattice, shown for several temperatures. The inset illustrates the first quadrant of the lattice. Here, and in all subsequent figures, when not shown, error bars are smaller than the symbol size.}
    \label{fig:Sb_cor}
\end{figure}

\section{Results}
\label{sec:results}

\subsection{Finite frequencies regime}
\label{subsec:adiabatic}

Motivated by the results from the bond-SSH model, we focus on the VBS phase and its potential competition with the AFM/CDW/SC one, for fixed $\omega_{0}=1$.
Then, we begin our analysis of the Hamiltonian in Eq.\,\eqref{Eq:SSHH} by examining the equal-time bond-bond correlation functions,
$$
C(|\mathbf{i}-\mathbf{j}|) = \frac{1}{2} \sum_{\hat{\alpha}=\hat{x}, \hat{y}}\sum_{\sigma, \sigma^{\prime}} \langle K_{\mathbf{i},\mathbf{i}+\hat{\alpha}}^\sigma K_{\mathbf{j},\mathbf{j}+\hat{\alpha}}^{\sigma^{\prime}} \rangle~,
$$
as the EPC strength increases\,\footnote{Throughout this section, we define $M=1$.}. Due to the $C_4$ symmetry of the lattice, here we average over the equivalent bonds, i.e.~along $Ox$ and $Oy$ directions.  Figure \ref{fig:Sb_cor} shows $C(|\mathbf{i}-\mathbf{j}|)$ along the high-symmetry points of the square lattice of linear size $L=12$ -- with the inset illustrating its first quadrant --, for different values of EPC, and fixed temperature \(T/t = (t\beta)^{-1} = 0.1\). As $\lambda/t$ increases, bond-bond correlations are significantly enhanced, indicating the emergence of a bond staggered pattern, similar to the bond-SSH model.
This feature suggests a staggered VBS ground state.

\begin{figure}[t]
     \includegraphics[scale=0.45]{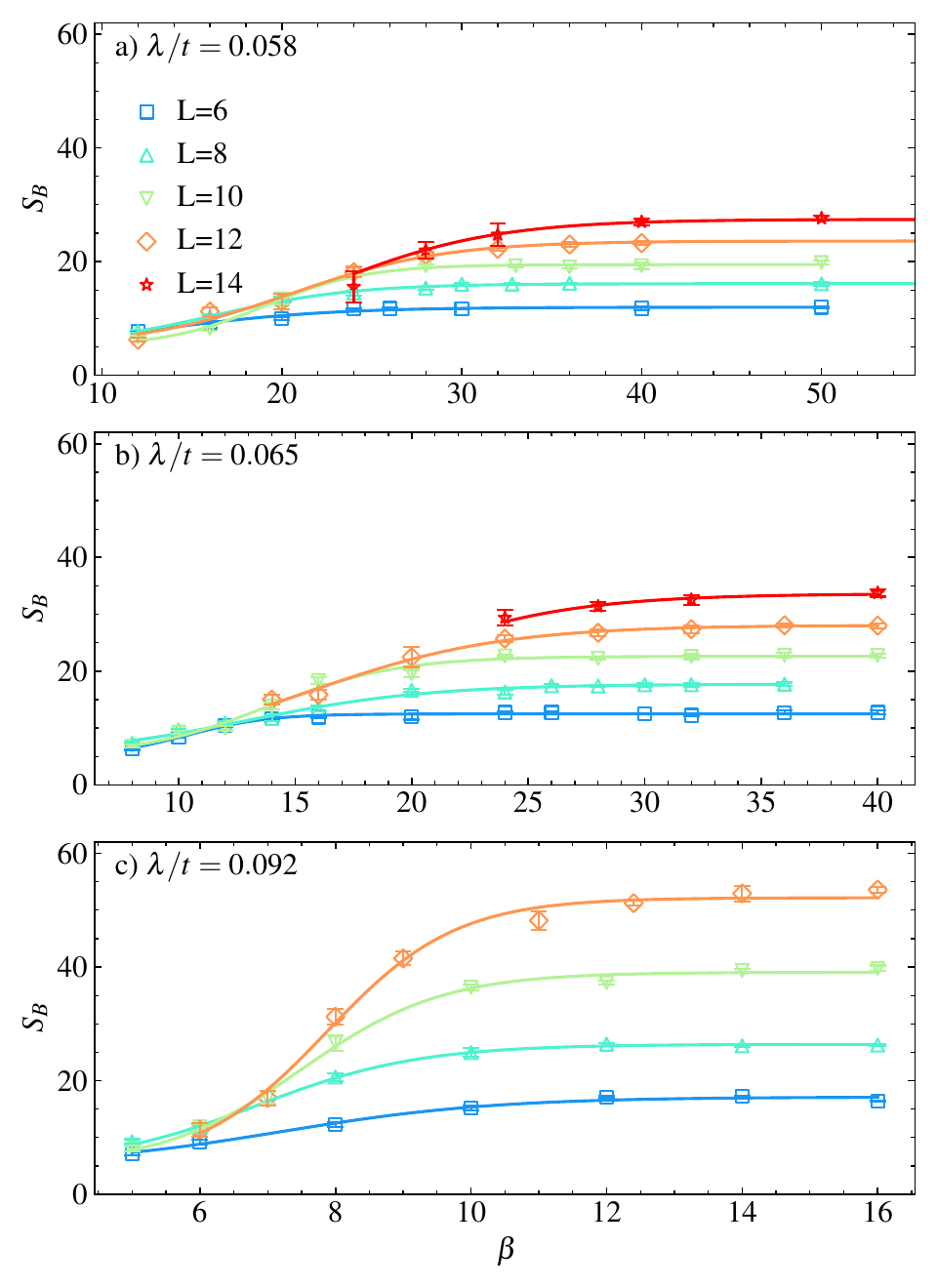}
     \caption{Staggered bond structure factor as a function of $\beta t$, for fixed $\omega_{0}=1$, and (a) $\lambda / t = 0.058$, (b) 0.065, and (c) 0.092, and several system sizes. The solid lines are just guides to the eye.}
     \label{fig:SBx}
\end{figure}

Given the large bond-bond correlations, we proceed analyzing the emergence of long-range order at low temperatures (i.e., in the ground state), leaving the analysis of temperature effects for later.
To this end, we examine the bond structure factors, Eqs. \eqref{Eq:SBx} and \eqref{Eq:SBy}. Due to the $C_4$ symmetry, we define $ S_{\rm B}(\mathbf{q}) =  S_{\rm B\,(x)} (\mathbf{q}) + S_{\rm B\,(y)}(\mathbf{q})$, which exhibits a peak at the wavevector $\mathbf{Q} = (\pi, \pi)$. This behavior corresponds to the four degenerate ground states of the staggered VBS phase. Figure \ref{fig:SBx} shows the behavior of the peak of the bond structure factor for fixed (a) $\lambda/t= 0.058$, (b) 0.065 and (c) 0.092 [or, equivalently, to $g/t =0.34$, 0.36, and 0.43, respectively].
Due to finite-size effects, the correlations stabilize into finite values as $\beta \to \infty$, which corresponds to the fact that the correlation length is larger than $L$.
As the coupling strength $\lambda/t$ increases, the stabilized values of $S_{\rm B}(\pi,\pi)$ also increase, with the size effects being even more pronounced, indicating the occurrence of a staggered VBS long-range order.

\begin{figure}[t]
    \includegraphics[scale= 0.45]{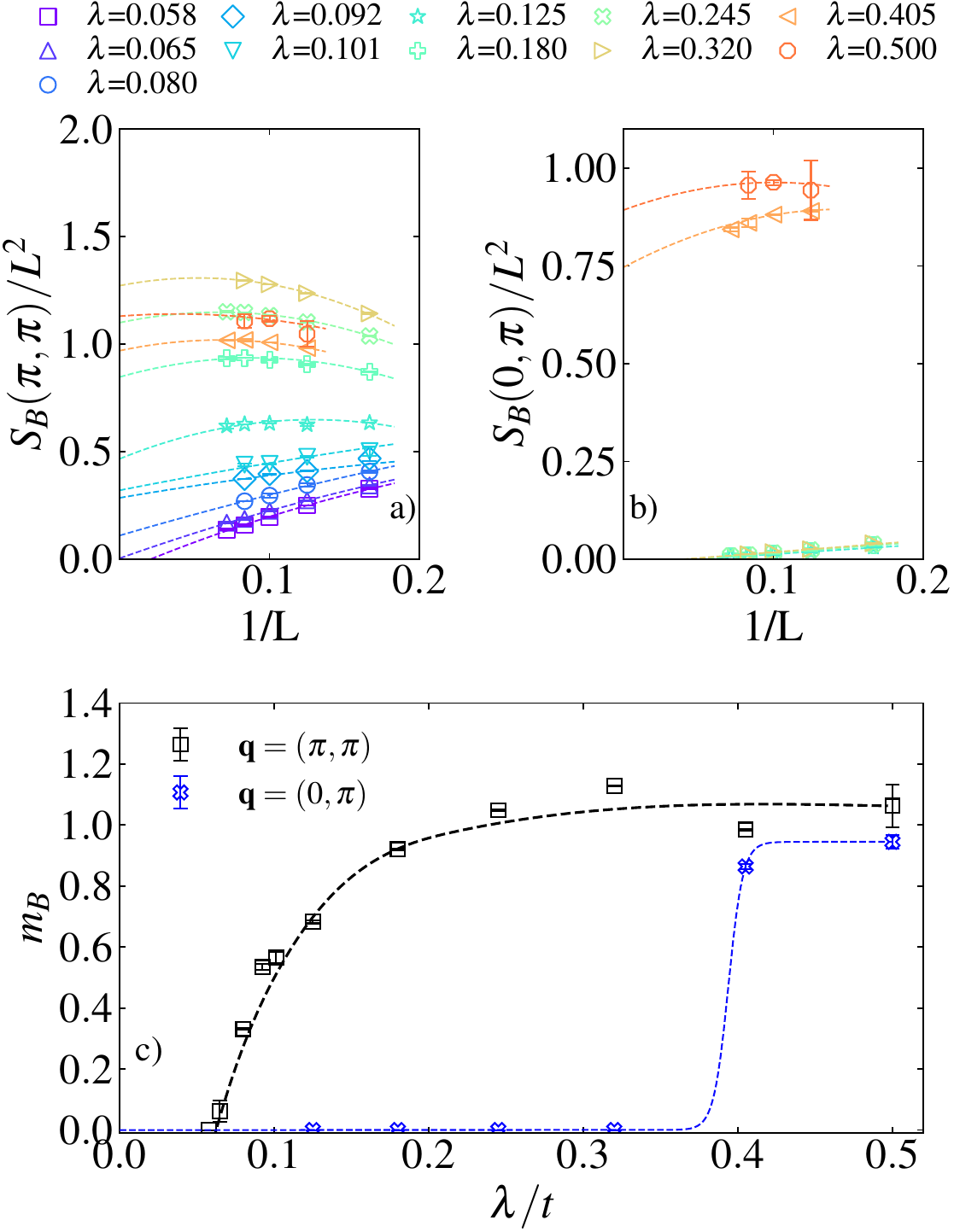}
    \caption{(a) Finite-size scaling analysis of the bond structure factors (a) $S_{\rm B} (\pi,\pi)$ and (b) $S_{\rm B} (0,\pi)$ at low temperatures, for different values of $\lambda$. (c) Extrapolated VBS order parameter as a function of $\lambda/t$, for both $(\pi,\pi)$ and $(0,\pi)$ modes. The dashed lines are guides to the eye.}
    \label{fig:fss}
\end{figure}

In order to determine the ground state VBS order parameter in the thermodynamic limit, $m_{\text{B}}$, we extrapolate the low-\textit{T} $S_{\rm B}(\pi,\pi)$ values for different lattice sizes, by performing
$$ \frac{S_{\text{B}}}{L^2} \equiv m^{2}_{\text{B}} + \frac{A}{L} + \mathcal{O}(1/L^2)~,$$
with \textit{A} being a constant determined in the fitting procedure.
Figure \ref{fig:fss}\,(a) shows the finite-size scaling of $S_{\rm B}(\pi,\pi)$ for various values of EPC, demonstrating the presence of a staggered VBS in the ground state for any $ \lambda \gtrsim \lambda_{c} \approx 0.058 t$ [or $g \gtrsim g_{c} \approx 0.34 t$].
The extrapolated values of the staggered $m_{\rm B}$ are then plotted in Fig.\,\ref{fig:fss}\,(c), in black square symbols.

\begin{figure}[t]
    \includegraphics[scale= 0.52]{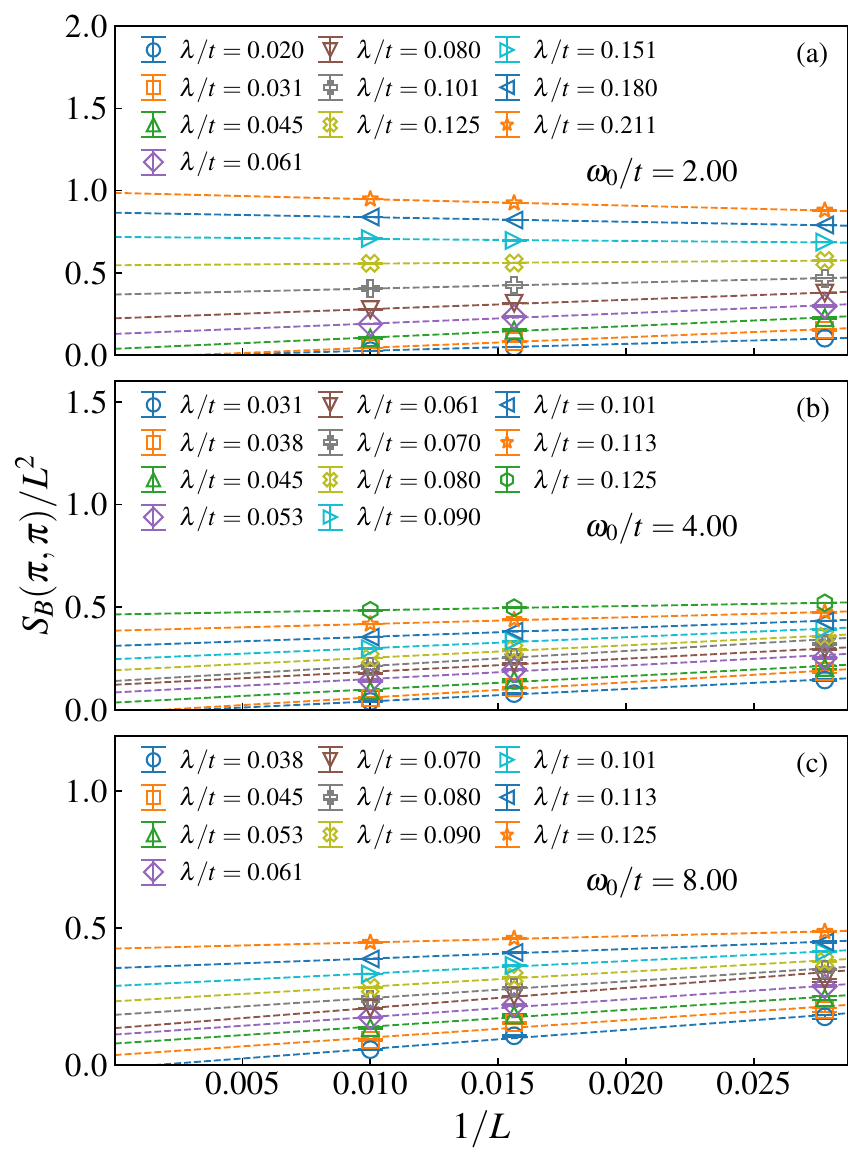}
    \caption{Finite-size scaling analysis of the bond structure factor $S_{\rm B} (\pi,\pi)$ for different values of $\lambda$, while fixing $\beta t = 40$,  (a) $\omega_{0}=2$, (b) $\omega_{0}=4$, and (c) $\omega_{0}=8$.}
    \label{fig:fsscritical}
\end{figure}

We repeated the same analysis for other frequencies, namely $\omega_{0}=2$, 4, and 8; the corresponding finite-size scaling results are shown in Fig.\,\ref{fig:fsscritical}\,(a)-(c). These data are consistent with a critical EPC in the range $\lambda_{c} \approx 0.045\text{--}0.050 t$, in line with the results for $\omega_{0}=1$ in Fig.\,\ref{fig:fss}\,(a).
The possible occurrence of a critical point at a finite $\lambda_{c}$ raises several issues. First, due to the nested Fermi surface and the van Hove singularity of the square lattice at half-filling, a metallic phase is inherently unstable for any interaction strength. Therefore, we expect that the system \textit{must} exhibit long-range order for any finite EPC, whether through an exponentially small bond order or through an AFM/CDW/SC phase.
Second, assuming that the suppression of the VBS structure factor is due to a critical point, the transition into the AFM/CDW/SC phase occurs at $\lambda/t$ values roughly an order of magnitude smaller than in the bond-SSH model\,\cite{goetz2021,Cai2021}. This comparison suggests that the dispersion in the electron-phonon interaction term of the optical-SSH model plays an important role in setting the critical scale.

\begin{figure}[t]
    \includegraphics[scale= 0.38]{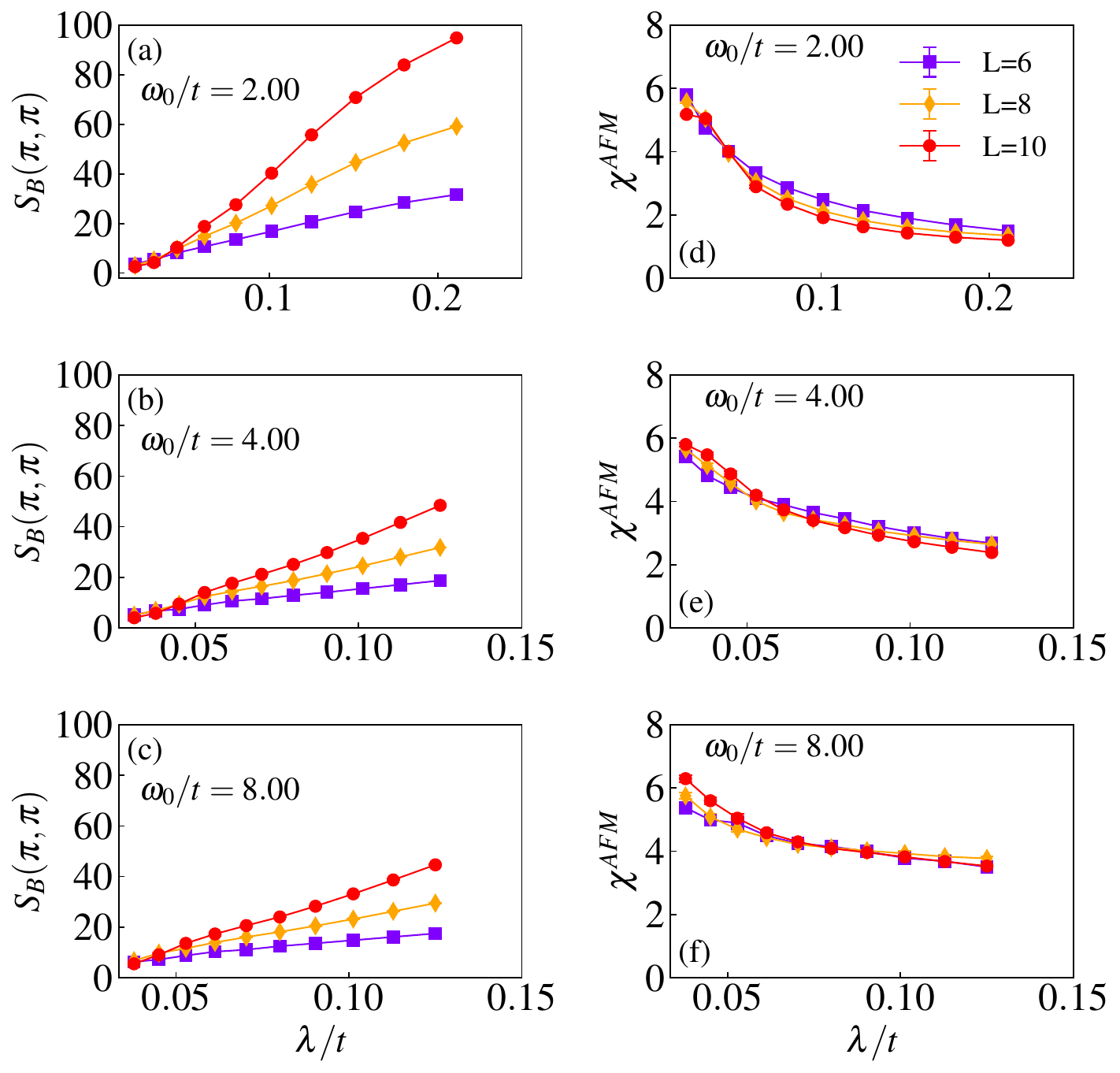}
    \caption{Bond structure factor $S_{\rm B} (\pi,\pi)$ as a function of $\lambda/t$, and different values of system sizes $L$, while fixing $\beta t = 40$,  (a) $\omega_{0}=2$, (b) $\omega_{0}=4$, and (c) $\omega_{0}=8$. Panels (d)-(f) show the corresponding results for the spin susceptibility.}
    \label{fig:SbSaf}
\end{figure}

In the limit $\lambda \to 0$, the staggered bond structure factor is strongly reduced, as shown in Fig.\,\ref{fig:SbSaf}\,(a)-(c) for the same set of parameters as in Fig.\,\ref{fig:fsscritical}. But, in contrast to the bond-SSH model, signatures of an AFM/CDW/SC phase are not evident here. To better capture spin-spin correlations, we analyze the staggered spin susceptibility,
\begin{equation}
\chi^{\mathrm{AFM}} = \frac{1}{N} \int_0^\beta d\tau \sum_{\mathbf{i},\mathbf{j}} e^{-i\mathbf{Q} \cdot (\mathbf{i} - \mathbf{j})} \langle S_\mathbf{i}^z(\tau) S_\mathbf{j}^z(0) \rangle,
\label{eq:xi_afm}
\end{equation}
shown in Fig.\,\ref{fig:SbSaf}\,(d)-(f).
Notice that $\chi^{\mathrm{AFM}}$ exhibits only a weak enhancement as $\lambda$ is reduced, which does not provide clear evidence for an AFM/CDW/SC phase in the ground state. In practice, approaching the $\beta \to \infty$ limit for $\lambda \to 0$ is challenging, and two scenarios remain viable within our numerical accuracy: (i) a VBS phase with an exponentially small order parameter, or (ii) a comparably small AFM/CDW/SC phase.
Other approaches to $\lambda \to 0$, such as fixing a large value for $g$ while increasing $\omega_{0}$, lead to similar results (not shown).

\begin{figure}[t]
    \includegraphics[scale =0.5]{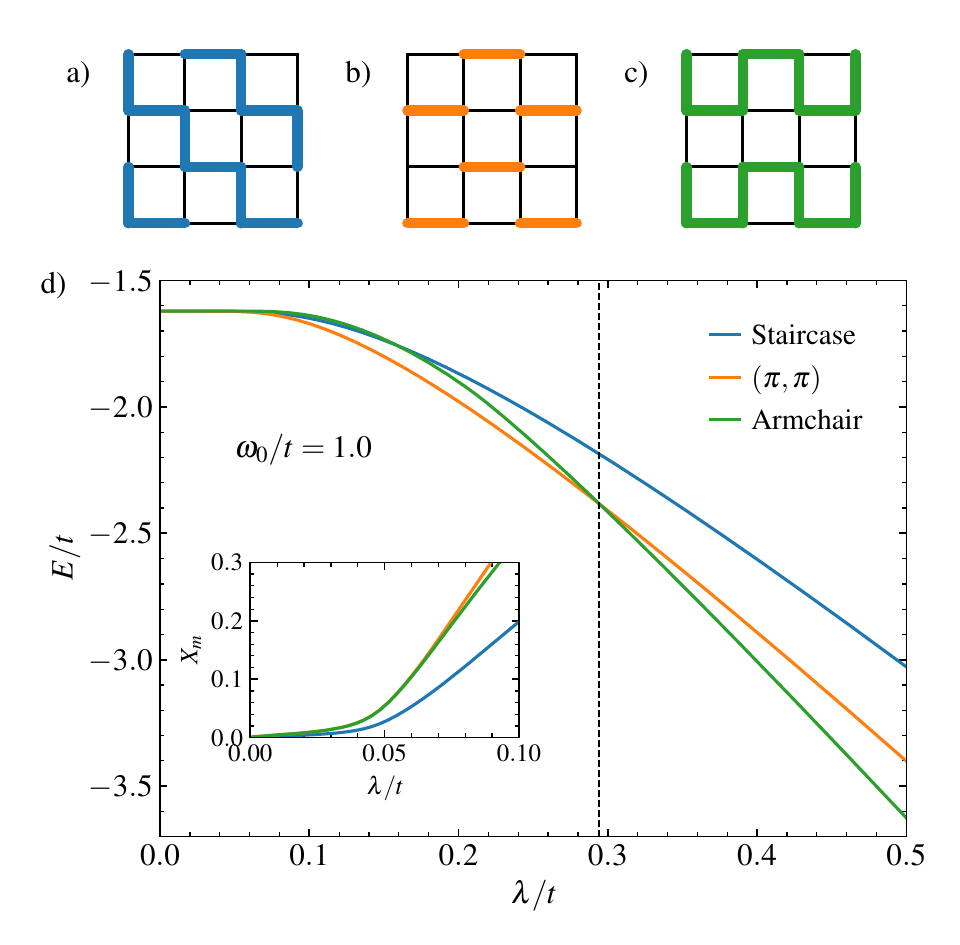}
    \caption{Patterns for (a) staircase, (b) staggered, and (c) armchair VBS configurations. Thick solid lines denotes stronger bonds, due to the permanent lattice deformation. (d) Mean-field internal energies of the adiabatic Hamiltonian for each VBS configuration as a function of $\lambda/t$. Inset: the expectation values for the permanent lattice deformations, $X_m = \langle Q_{\mathbf{i}, \alpha} \rangle$.}
    \label{fig:ansatz}
\end{figure}

In view of this, we examine two limits of the model to probe a possible VBS-to-AFM/CDW/SC transition: (i) $\omega_{0} \to 0$ and (ii) $\omega_{0}\to\infty$. As discussed in Sec.\,\ref{sec:model}, the former corresponds to static lattice distortions, where phonon dynamics are neglected, and translational symmetry may be broken. Therefore, this limit can be examined by the mean-field Hamiltonian in Eq.\,\eqref{Eq:MFTSSHH}. We use three different ans\"atze for the distortion pattern: (i) staggered, (ii) armchair, and (iii) stair-like VBS configurations, as illustrated in Figs.\,\ref{fig:ansatz}\,(a)-(c), respectively.
The mean-field ground state energies are shown in Fig.\,\ref{fig:ansatz}\,(d) for the three considered cases. For all $\lambda > 0$, the solution corresponds to a VBS phase, in particular being the staggered one in the limit $\lambda \to 0$, with the order parameter being exponentially small, as shown in the inset.
Indeed, these results do not support a metallic ground state in the weak coupling limit, as expected; also, it may indirectly support a weak VBS scenario at finite $\omega_{0}$, when and $\lambda \to 0$.
We return to this point in our analysis of the anti-adiabatic limit.

Interestingly, within the mean-field description, increasing $\lambda$ leads to a change from the VBS pattern, with a transition from the staggered to the armchair configuration, a behavior not expected in the bond-SSH model.
As the MFT solution is exact in the limit $\omega_{0} \to 0$, it would be relevant to investigate the behavior of the VBS phase for large values of EPC, within the QMC approach.
Given this, and returning to the QMC analysis, we employ simulations for larger values of $\lambda$, and fixed $\omega_{0} = 1$. As result, we find that, at $\lambda/t \approx 0.37$, the $\mathbf{q} = (0,\pi)$ mode of $S_{\rm B}(\mathbf{q})$ is enhanced, in addition to the $\mathbf{q} = (\pi,\pi)$ mode, in line with the MFT expectations. The extrapolation of $S_{\rm B}(\mathbf{q})$ as a function of $1/L$ indicates that the $\mathbf{q} = (0,\pi)$ peak persists in the thermodynamic limit, as shown in Fig.\,\ref{fig:fss}\,(b). The appearance of this columnar mode is accompanied by a reduction in the staggered mode, as displayed in Fig.\,\ref{fig:fss}\,(c), leading to an armchair mode with inhomogeneous bonds.

\begin{figure}[t]
    \includegraphics[scale=0.5]{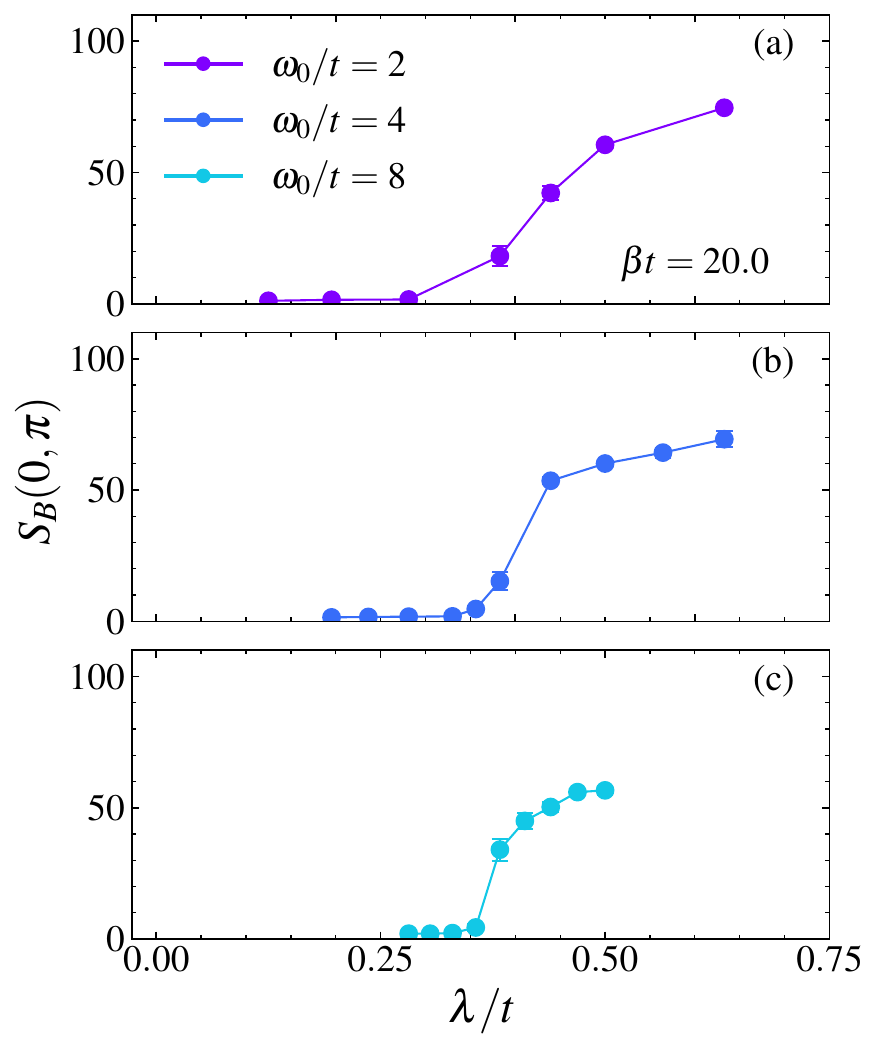}
\caption{Columnar bond structure factor $S_{B}(0,\pi)$ as a function of $\lambda/t$ ($\equiv  \frac{g^2}{2\omega_{0}^2}$), and fixed (a) $\omega_{0} = 2$, (b) 4 and (c) 8. The calculations were performed for fixed $\beta t=20$, and $L=8$.}
\label{fig:Sb_0pi_omega}
\end{figure}

The staggered-to-armchair VBS phase transition in the ground state is observed even for large values of $\omega_{0}$. Figure \ref{fig:Sb_0pi_omega} shows the behavior of $S_{\rm B}(0,\pi)$, for fixed $\beta t = 20$ and $L = 8$. As $\omega_{0}$ increases, the columnar mode remains present for a given value of $\lambda$, but with reduced amplitude. This weakness of the bond–bond correlations suggests that higher phonon frequencies favor spin-spin correlations, although the ground state remains in either the staggered or the armchair VBS phase for large EPC.

It is important to mention that the staggered VBS phase is characterized by ordering at wavevector $\mathbf{q} = (\pi,\pi)$, which breaks lattice translation symmetry down to translations by $\mathbf{A}^{\text{Stag}}_{\pm} = \mathbf{a}_1 \pm \mathbf{a}_2$ and results in four degenerate ground states. Similarly, the armchair order further reduces translation symmetry, remaining invariant only under translations by $\mathbf{A}^{\text{Arm}}_{+} = 2\mathbf{a}_1$ and $\mathbf{A}^{\text{Arm}}_{-} = 2\mathbf{a}_2$, and is consequently eightfold degenerate.
We emphasize that the emergence of the columnar mode (that leads to the armchair phase) occurs within the staggered phase, so the eightfold degeneracy may be understood as: for each of the four staggered configurations, there are  two possible columnar states.
This implies that the transition from the staggered to the armchair phase corresponds to the breaking of an \textit{additional} discrete $\mathbb{Z}_2$ symmetry. Therefore, the associated order parameter is Ising-like, and the transition is expected to belong to the three-dimensional Ising universality class, with a critical exponent $\beta \simeq 0.3$ [see, e.g., Ref.\,\onlinecite{GARCIA2003}, and references therein].
The relatively small exponent can make the staggered-to-armchair VBS phase transition appear weakly first-order in numerical simulations, despite being continuous.

\subsection{The antiadiabatic limit}

\begin{figure}[t]
    \includegraphics[scale=0.38]{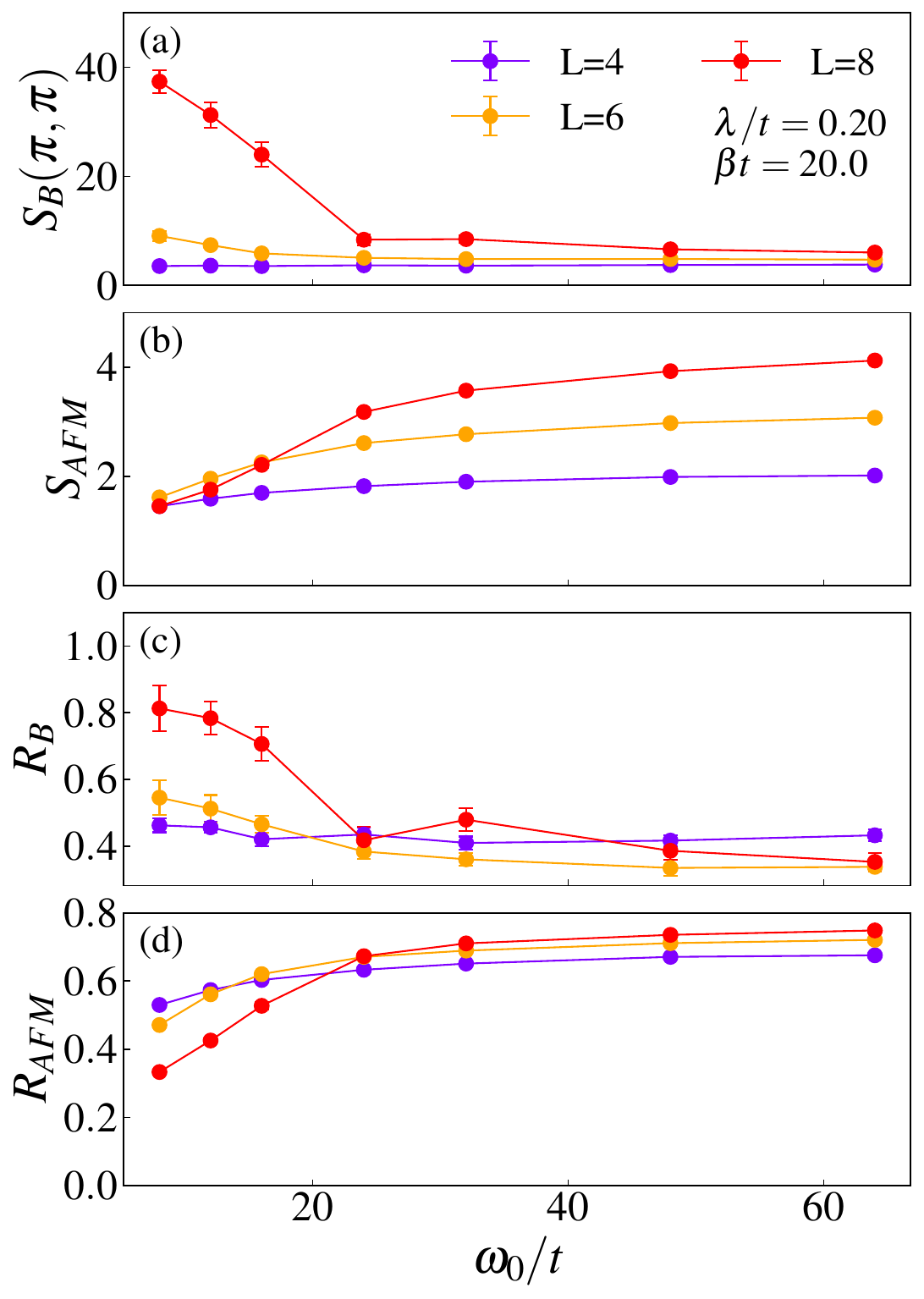}
\caption{ (a) Bond structure factor $S_{\rm B} (\pi,\pi)$ and (b) staggered spin structure factor $S_{\rm AFM}$, and their corresponding correlation ratios (c) $R_{\rm B}$ and (d) $R_{\rm AFM}$ as functions of $\omega_{0}$, for different values of $L$, while fixing $\beta t = 20$ and $\lambda = 0.20t$.}
\label{fig:omegavar}
\end{figure}

Before analyzing the true $\omega_{0} \to \infty$ limit, described in the Hamiltonian of Eq.\,\eqref{eq:SSH_antiadiabatic}, it is useful to examine how the responses of the finite frequencies Hamiltonian in Eq.\,\eqref{Eq:SSHH} evolve with increasing $\omega_{0}$ at fixed EPC. Figures \ref{fig:omegavar}\,(a) show the bond $S_{\mathrm{B}}(\pi,\pi)$ and (b) the staggered spin $S_{\mathrm{AFM}}$ structure factors as functions of $\omega_{0}$, and fixed $\beta t=20$, $\lambda=0.20t$, and several system sizes. For $\omega_{0} \approx 20$, $S_{\mathrm{B}}$ is substantially suppressed, while $S_{\mathrm{AFM}}$ increases, with the trend visible across the sizes shown. In contrast to the $\lambda \to 0$ limit discussed earlier, these data support antiferromagnetic correlations consistent with AFM long-range order at large $\omega_{0}$.

A further probe of the AFM phase may be provided by the bond and spin correlation ratios
\begin{equation}
R_{\alpha}(L) = 1 - \frac{S_{\alpha}(\mathbf{Q} - \delta \mathbf{q})}{S_{\alpha}(\mathbf{Q})},
\label{eq:Rc_bond}
\end{equation}
with the index $\alpha=\mathrm{B}$ or $\mathrm{AFM}$ denoting the bond and staggered spin structure factors, respectively\,\cite{Kaul2015,Sato2018,Liu2018}. Figures \ref{fig:omegavar}\,(c) and (d) show $R_{\mathrm{B}}(L)$ and $R_{\mathrm{AFM}}(L)$ as functions of $\omega_{0}$ at fixed $\lambda=0.20 t$, $\mathbf{Q}=(\pi,\pi)$, and $|\delta \mathbf{q}| = 2\pi/L$. Consistent with the preceding analysis, the correlation ratios display crossings near $\omega_{0}\approx 20$, which is the first clear evidence for the emergence of long-range order AFM/CDW/SC in the anti-adiabatic limit. At this point, we note that the transition to the AFM phase occurs at frequencies much larger than in the bond-SSH model, which emphasizes the stronger VBS phase in the optical model \footnote{Because dimerization is energetically favored in the optical-SSH model, the integrated autocorrelation times are substantially larger than in the bond-SSH case. For large EPC, the simulations typically need to be extended by on the order of $10^{5}$ Monte Carlo sweeps to achieve comparable statistical accuracy.}. We note that the emergence of the AFM/CDW/SC order at large phonon frequencies and strong EPC does not, by itself, establish adiabatic continuity with the non-VBS state obtained for $\lambda \to 0$. Additional analysis of the anti-adiabatic limit is required to determine any connection.

Now we turn to investigate the $\omega_{0} \to \infty$ limit. We recall that, for the bond-SSH model in the high-frequency regime, the EPC term becomes
$$
\left( K^{\uparrow}_{\mathbf{i},\mathbf{j}} + K^{\downarrow}_{\mathbf{i},\mathbf{j}} \right)^2 \propto 4 \left( \mathbf{S}_\mathbf{i} \cdot \mathbf{S}_\mathbf{j} + \boldsymbol{\eta}_\mathbf{i} \cdot \boldsymbol{\eta}_\mathbf{j} \right),
$$
where $\mathbf S_{\mathbf i}$ denotes the spin operator, associated with AFM order, and $\boldsymbol{\eta}_i$ is the Anderson pseudospin operator, associated with $s$-wave SC and CDW orders\,\cite{goetz2021,Cai2021}.
Within this description, increasing the phonon frequency $\omega_0$ should lead to a phase transition from a VBS phase into an AFM/CDW/SC one.
For the optical-SSH model, the situation is much less clear. Indeed, in the limit $\omega_0 \to \infty$, the effective interaction term
$\Big[ \sum_{\sigma} \Big(K_{\mathbf{i},\mathbf{i}+\hat{\alpha}}^{\sigma} - K_{\mathbf{i}-\hat{\alpha},\mathbf{i}}^{\sigma}\Big)\Big]^2$ in Eq.\,\eqref{eq:SSH_antiadiabatic}
energetically favors dimerization, particularly in weak coupling regime, providing a simple mechanism to lower the energy when two neighboring bonds are different. This should lead to a strong competition between VBS and other possible orders, so that a careful analysis of this limit is required to determine the nature of the ground state.

\begin{figure}[t]
    \includegraphics[scale=1.20]{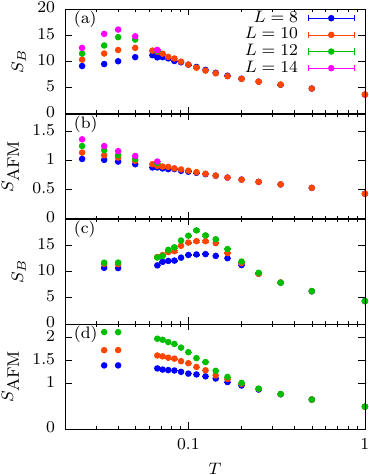}
\caption{(a) Bond structure factor $S_{\rm B}(\pi,\pi)$ and (b) staggered spin structure factor $S_{\rm AFM}$ as functions of the temperature, for fixed $\lambda=0.16t$, and several system sizes for the anti-adiabatic Hamiltonian $\mathcal{H}_{\infty}$ of Eq.\,\eqref{eq:SSH_antiadiabatic}. The same for panels (c) and (d), but for fixed $\lambda=0.25t$.}
\label{fig:Sb_Sssc_antiadiabatic}
\end{figure}

We start our analysis of the effective Hamiltonian $\mathcal{H}_{\infty}$ in Eq.\,\eqref{eq:SSH_antiadiabatic} by examining the bond and spin correlations. Figures \ref{fig:Sb_Sssc_antiadiabatic}(a)  and \ref{fig:Sb_Sssc_antiadiabatic}(b) respectively show the bond structure factor $S_{\mathrm B}(\pi,\pi)$ and the staggered spin structure factor $S_{\mathrm{AFM}}$ as functions of temperature, for fixed $\lambda=0.15 t$, and for several linear sizes $L$. As $T\to 0$, $S_{\mathrm B}(\pi,\pi)$ increases and exhibits a finite-size dependence; it also develops a maximum before decreasing. By contrast, $S_{\mathrm{AFM}}$ remains small but exhibits a weak dependence on $L$ at the lowest temperatures. These responses show a strong competition between bond and spin correlations, with the former being more robust at finite $T$. However, the peak in $S_{\mathrm B}(\pi,\pi)$ suggests that bond correlations tend to saturate or weaken at lower temperatures, while the spin channel may become more relevant upon further cooling.
Indeed, this is confirmed by increasing the EPC to $\lambda=0.25 t$, as presented in Fig.\,\ref{fig:Sb_Sssc_antiadiabatic} (c) and (d) for the $S_{\mathrm B}(\pi,\pi)$ and $S_{\mathrm{AFM}}$, respectively.
Upon cooling, $S_{\mathrm B}(\pi,\pi)$ decreases and saturates in a large, but weakly size-dependent value. By contrast, the staggered spin structure factor increases with decreasing temperature, indicating a strengthening of antiferromagnetic correlations, and showing long-range order (not shown).

\begin{figure}[t]
\includegraphics[scale=1.5]{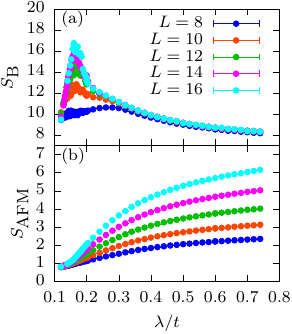}
\caption{(a) Bond structure factor $S_{\rm B}(\pi,\pi)$ and (b) staggered spin structure factor $S_{\rm AFM}$ as functions of $\lambda$, for fixed inverse temperature $\beta t=25$, and several system sizes for the anti-adiabatic Hamiltonian $\mathcal{H}_{\infty}$ of Eq.\,\eqref{eq:SSH_antiadiabatic}.}
\label{fig:S_lambda}
\end{figure}

The competition between bond and spin correlations is more evident when the temperature is fixed at a low value and the EPC is varied. Figures \ref{fig:S_lambda}(a) and \ref{fig:S_lambda}(b) respectively show the staggered bond and spin structure factors, $S_{\rm B}$ and $S_{\rm AFM}$, as functions of $\lambda$ at fixed $\beta t=25$. At large EPC, spin correlations dominate; at weak coupling, enhanced bond correlations are observed, evidenced by a peak in $S_{\rm B}$ (its rapid suppression as $\lambda \to 0$ is a finite temperature effect). Because $\lambda=0$ is a singular point that favors both VBS and AFM phases, the data in Fig.\,\ref{fig:S_lambda} might suggest a narrow region with VBS order near $\lambda \approx 0$. However, this is unlikely. First, despite the enhancement of $S_{\rm B}$ at weak coupling, finite-size scaling of $S_{\rm B}/L^2$ does not yield a nonzero thermodynamic order parameter, even at the peak of $S_{\rm B}$. Second, further lowering the temperature does not increase the peak height, but shifts it to smaller $\lambda$, as shown in Fig.\,\ref{fig:Sb_Sssc_antiadiabatic}. Third, if both VBS and AFM instabilities were present as $\lambda \to 0$, the continuous-symmetry AFM state would be expected to be favored over the discrete VBS, due to the presence of Goldstone modes. Therefore, these observations do not support a VBS phase in the antiadiabatic limit.

\begin{figure}[t]
\includegraphics[scale=1.5]{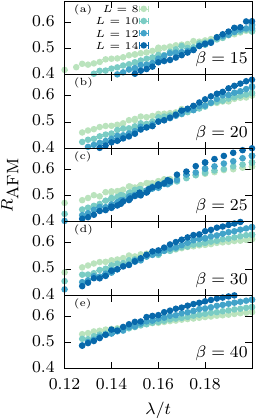}
\caption{Spin correlation ratio $R_{\rm AFM}$ as a function of $\lambda$, for different system sizes and fixed (a) $\beta t = 15$, (b) $\beta t = 20$, (c) $\beta t = 25$, (d) $\beta t = 30$, and (e) $\beta t = 40$ for the anti-adiabatic Hamiltonian $\mathcal{H}_{\infty}$ of Eq.\,\eqref{eq:SSH_antiadiabatic}.}
\label{fig:RK_Rssc_antiadiabatic}
\end{figure}

In order to further probe long-range AFM order, we compute the spin correlation ratio [Eq.\,\eqref{eq:Rc_bond}]. Figure \ref{fig:RK_Rssc_antiadiabatic} (a)-(e) shows $R_{\rm AFM}$ as a function of $\lambda$ at fixed $\beta t = 15$, 20, 25, 30, and 40, respectively, for several system sizes $L$. For each $\beta$, there are crossings in $R_{\rm AFM}(L)$ consistent with a continuous transition. However, as $\beta$ increases, these crossings shift to smaller $\lambda$, indicating a finite critical coupling $\lambda_c$ is an artifact of temperature effects. In other words, we expect that $\lambda_c \to 0$ as $\beta \to \infty$, with the ground-state exhibiting an AFM/CDW/SC phase for any finite EPC.

Taken together with the QMC and MFT results of the preceding subsection, our data provide the ground state phase diagram in Fig.\,\ref{fig:phase_diagram}. We emphasize that, in the antiadiabatic limit $\omega_{0} \to \infty$, an AFM/CDW/SC phase appears for any $\lambda>0$. Hence, an adiabatic connection from this phase into the non-VBS state at weak EPC and finite $\omega_{0}$ is possible, strongly suggesting that this region likewise exhibits AFM/CDW/SC order.

\subsection{Critical temperatures}

\begin{figure}[t]
\includegraphics[scale=0.5]{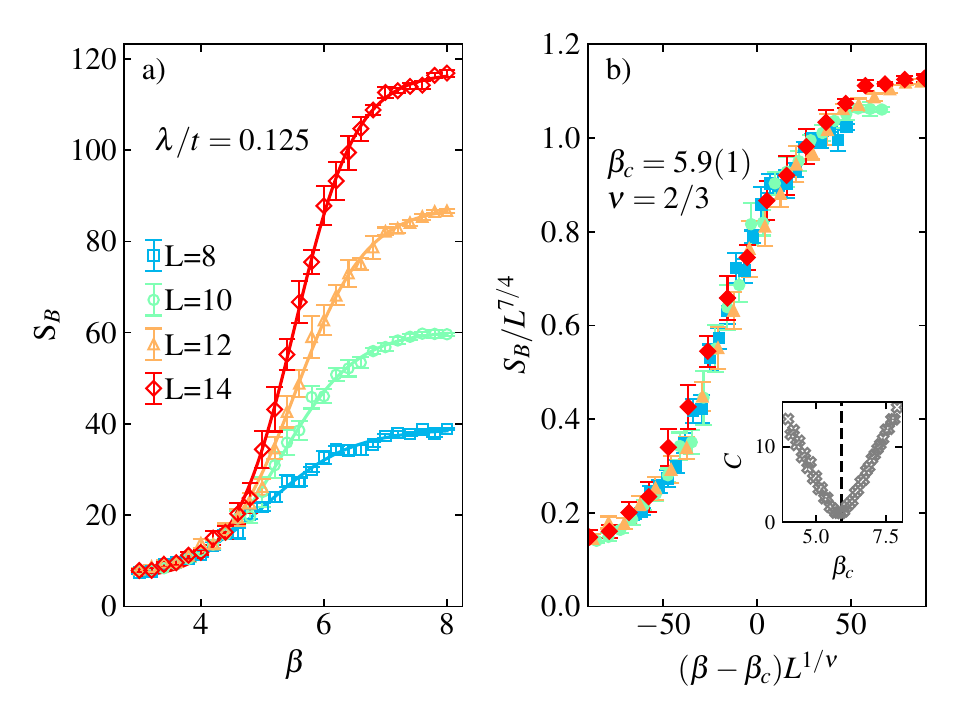}
\caption{Data collapse resulting from the scaling analysis of the bond Structure factor. Inset: Cost function as the critical value of $\beta$ is varied, from which we obtain the values: $\beta_c=5.9 \pm 0.1$}
 \label{fig:Sbcollapse}
\end{figure}

Having established the ground state of the optical-SSH model, we now examine the critical temperatures for the emergence of a VBS phase.  
To this end, we examine the staggered bond structure factor $S_{\rm B}(\pi,\pi)$ as a function of the inverse of temperature for fixed $\lambda=0.125 t$ and $\omega_0=1$, as shown in Fig.\,\ref{fig:Sbcollapse}\,(a). Notice that, as $\beta$ increases, $S_{\rm B}(\pi,\pi)$ grows and develops a noticeable size dependence at larger $\beta$. The onset of this size dependence defines a characteristic energy scale, which, in the thermodynamic limit, would be consistent with the critical $\beta_c$.
Indeed, the staggered VBS admits bond orientation along $Ox$ or $Oy$, yielding four equivalent ground states. The ordered phase therefore breaks a fourfold discrete $Z_4$ symmetry of the square lattice, which in two dimensions allows for a finite-$T$ transition.

However, critical behavior in $Z_4$-broken systems is subtle. For instance, in the 2D Ashkin-Teller model, the critical exponents may vary continuously from those of the 4-state Potts model to the Ising model\,\cite{songbo2012,Songbo2013}. Therefore, and to simplify the following discussions, here we assume that our VBS finite temperature phase transitions belong to the same universality class as the Potts model with $q=4$\,\cite{Wu1982}.
We validate this assumption by employing a data collapse for
\begin{equation}
\frac{S_{\mathrm{B}}}{L^{\gamma/\nu}} = f\left[(\beta - \beta_c) L^{1/\nu}\right]~,
\label{eq:scaling_VBS}
\end{equation}
fixing the critical exponents $\gamma = 7/6$ and $\nu = 2/3$, whose results are displayed in Fig.\,\ref{fig:Sbcollapse}\,(b).
The inset shows the minimization of the cost function $C(\beta)$, defined in Ref.\,\onlinecite{Suntajs2020}, from which we estimate the critical inverse temperature $\beta_c t = 5.9(1)$.

\begin{figure}[t]
     \includegraphics[scale=0.5]{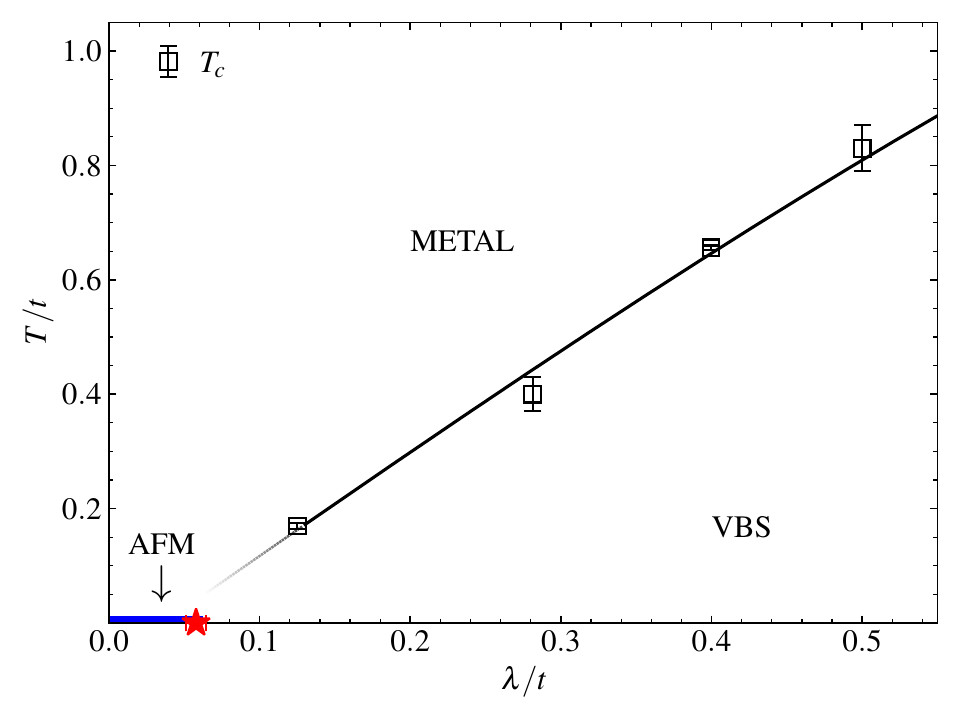}
     \caption{Finite temperature phase diagram. The black squares represent the VBS critical temperature. The red star indicates the ground state critical point. }
     \label{fig:diagramT}
 \end{figure}

Repeating the scaling collapse at other electron-phonon couplings, we obtain $T_{c}(\lambda)$; the resulting critical temperatures are compiled in the finite temperature phase diagram in Fig.\,\ref{fig:diagramT}.
Within the range of interactions analyzed here, the critical temperatures reach values as high as $T_{\mathrm{c}} \sim 0.83 t$ for $\lambda = 0.5 t$. Also, we do not observe a maximum in the VBS transition temperature, which suggests that $T_{c}$ can be even larger.

At this point, it is worth recalling the features of the Holstein model, where the phonons couple to the electronic density, and a CDW phase emerges at the ground state for any $\lambda >0$\,\cite{Costa2020}.
For this model, the CDW critical temperature $T_{\mathrm{CDW}}$ rises with increasing EPC in the weak and intermediate regimes. This can be understood from a mean-field point of view, since stronger coupling enhances the phonon-mediated attraction between electrons. However, as the coupling continues to grow, electrons become heavily dressed by local lattice distortions, and their effective mass increases exponentially ($m^* \sim e^{g^2/\omega_0^2}$), which results in a drastic suppression of carrier mobility in the metallic phase at high temperatures\,\cite{Lang1964,Bonca1999,Bonca2000}. Since long-range order relies on collective motion, this electron mass renormalization reduces the effective hopping, which, in turn, alters the transition temperature, producing a maximum in $T_{\mathrm{CDW}}$ at intermediate coupling. Indeed, such behavior occurs despite the details of lattice geometry or electronic band dispersion, being a property of the type of electron-phonon coupling\,\cite{Feng2020,CohenStead2020,Zhang20192}. For comparison, the largest CDW transition temperature in the Holstein model is $T_{\mathrm{CDW}}\approx 0.26 t$ for the square lattice\,\cite{Feng2020} and $T_{\mathrm{CDW}}\approx 0.40 t$ for the cubic lattice\,\cite{CohenStead2020}. Therefore, when $\lambda \to \infty$, $T_{\mathrm{CDW}}\to 0 $, and exotic phases may appear, such as a Bose insulating phase\,\cite{Xiao2021}. 

By contrast, when the EPC acts modulating the hopping amplitude, as in the SSH model, the physical responses change qualitatively.  As discussed in Refs.\,\onlinecite{marchand2010,Sous2018}, the SSH model leads to light polarons and bipolarons, even at strong coupling. That is, the effective mass does not diverge as in the Holstein model. This fundamental difference implies that, in many-body SSH-type systems, ordered states arising from bond distortions may have higher characteristic ordering temperatures. Indeed, this is confirmed by our QMC analysis of the critical temperatures of the VBS phase, which are much higher than those of the Holstein model.
In line with our claim, the emergence of pairing away from half-filling at high critical temperatures was recently predicted for the bond-SSH model\,\cite{cai2023}. 
As a final comment, we note that, in contrast to the bond-SSH model, $\langle Q_{\alpha,\mathbf i}\rangle \approx 0$ for the optical-SSH model, even at strong EPC (not shown). This reduces the phonon energy cost and favors long-range orders driven by the electron-phonon term. On simple grounds, we therefore expect a larger pairing response in doped systems for the optical-SSH model, although a quantitative analysis is beyond the scope of this work.

\section{Conclusions}
\label{sec:conclusions}

In this work, we have investigated some ground state and finite temperature properties of the two-dimensional optical Su-Schrieffer-Heeger model, emphasizing the differences and similarities with its bond-type counterpart and other electron-phonon models, such as the Holstein model. To this end, we employed sign-problem-free auxiliary-field quantum Monte Carlo simulations and complementary mean-field analyses, having established ground state and finite temperature phase diagrams.

For any finite frequency, our calculations found a staggered VBS for arbitrarily weak EPC. Increasing the coupling drives a transition to an armchair (staggered + columnar) VBS phase, consistent with a 3D Ising universality class. However, for finite $\omega_{0}$, the staggered VBS appears only beyond a critical coupling $\lambda_{c}$, while the non-VBS region has enhanced spin correlations. Upon further increase in $\omega_{0}$, the simulations indicate that, for sufficiently large phonon frequencies, the system exhibits a transition from VBS to AFM/CDW/SC phases, which suggests that the non-VBS region at weak $\lambda$ is also AFM/CDW/SC. Here, we emphasize that the critical couplings and frequencies for such a transition differ substantially from those of the bond-SSH model. For example, in the optical-SSH case, one needs $\omega_{0} \approx 20$ to reach the transition, one order of magnitude larger than the bond-case.

In the antiadiabatic limit, when $\omega_{0} \to \infty$, integrating out the phonons yields an effective fermionic Hamiltonian that makes explicit the competition between bond and spin correlations. Indeed, the analysis of this Hamiltonian indicates a strong VBS phase at weak coupling, but with the ground state exhibiting AFM/CDW/SC order.

Our results provide a comprehensive characterization of the optical-SSH model, showing how its intrinsic phonon dispersion qualitatively alters the balance between competing ordered phases in both zero and finite temperature cases.
This study opens the opportunity to further understand some extensions, such as adding other interactions or doping the system, which may connect to experimental realizations in low-dimensional quantum materials.

\section*{ACKNOWLEDGMENTS}

The authors are grateful to S.~Sorella, F.~Becca, D. Piccioni, and R.R.~dos Santos for the many insightful discussions on the subject.
N.C.C., S.A.S.J., and J.L.P.S thank the Brazilian Agencies Conselho Nacional de Desenvolvimento Cient\'ifico e Tecnol\'ogico (CNPq), Coordena\c c\~ao de Aperfei\c coamento de Pessoal de Ensino Superior (CAPES), and Fundação de Amparo \`a Pesquisa do Estado do Rio de Janeiro, FAPERJ.
S.A.S.J.~thanks CNPq, grant No.~201000/2024-5.
N.C.C.~acknowledges support from FAPERJ Grants No.~E-26/200.258/2023 [SEI-260003/000623/2023] and E-26/210.592/2025 [SEI-260003/004500/2025], CNPq Grants No.~313065/2021-7 and 308130/2025-1, and Serrapilheira Institute Grant No.~R-2502-52037, and Alexander von Humboldt Foundation.
FA~thanks the W\"urzburg-Dresden Cluster of Excellence on Complexity and Topology in Quantum Matter ct.qmat (EXC 2147, project-id 390858490).   G.R.~thanks the DFG for financial support under Grant No. AS 120/19-1 (Project No. 530989922).
The authors gratefully acknowledge the Gauss Centre for Supercomputing e.V. (www.gauss-centre.eu) for funding this project by providing computing time on the GCS Supercomputer SUPERMUC-NG at Leibniz Supercomputing Centre
(www.lrz.de).  We  also gratefully acknowledge the scientific support and HPC resources provided by  the Erlangen National High Performance Computing Center (NHR@FAU) of the Friedrich-Alexander-Universität Erlangen-Nürnberg (FAU) under NHR project 80069 provided by federal and Bavarian state authorities. NHR@FAU hardware is partially funded by the German Research Foundation (DFG) through grant 440719683.

\bibliography{refe.bib}

\end{document}